\documentclass[%
reprint,
showpacs,preprintnumbers,
nofootinbib,
amsmath,amssymb,
aps,
prd,
floatfix,
]{revtex4-1}

\usepackage{graphicx}% Include figure files
\usepackage{dcolumn}% Align table columns on decimal point
\usepackage{bm}% bold math

\usepackage{url}

\newcommand{\boss}[2]{\ensuremath{\rlap{\kern-2.5pt\ensuremath{\overset{\scriptscriptstyle(-)}{\phantom{#1}}}}{\ensuremath{{#1}_{#2}}}}}

\begin{document}

\preprint{\begin{tabular}{l}
\texttt{EURONU-WP6-10-30}
\\
\texttt{arXiv:1012.0267v3 [hep-ph]}
\\
\texttt{Phys.Rev.D 83 (2011) 053006}
\end{tabular}}

\title{Large Short-Baseline $\bar\nu_{\mu}$ Disappearance}

\author{Carlo Giunti}
\email{giunti@to.infn.it}
\altaffiliation[also at ]{Department of Theoretical Physics, University of Torino, Italy}
\affiliation{INFN, Sezione di Torino, Via P. Giuria 1, I--10125 Torino, Italy}

\author{Marco Laveder}
\email{laveder@pd.infn.it}
\affiliation{Dipartimento di Fisica ``G. Galilei'', Universit\`a di Padova,
and
INFN, Sezione di Padova,
Via F. Marzolo 8, I--35131 Padova, Italy}

\date{\today}

\begin{abstract}
We analyze
the LSND, KARMEN and MiniBooNE data on
short-baseline
$\bar\nu_{\mu}\to\bar\nu_{e}$ oscillations and
the data on short-baseline
$\bar\nu_{e}$ disappearance
obtained in the Bugey-3 and CHOOZ reactor experiments
in the framework of 3+1 antineutrino mixing,
taking into account
the MINOS observation of long-baseline $\bar\nu_{\mu}$ disappearance
and
the KamLAND observation of very-long-baseline $\bar\nu_{e}$ disappearance.
We show that the fit of the data implies
that the
short-baseline disappearance of $\bar\nu_{\mu}$
is relatively large.
We obtain a prediction of an effective amplitude
$\sin^2 2\vartheta_{\mu\mu} \gtrsim 0.1$
for short-baseline $\bar\nu_{\mu}$ disappearance
generated by
$0.2 \lesssim \Delta{m}^2 \lesssim 1 \, \text{eV}^2$,
which could be measured in future experiments.
\end{abstract}

\pacs{14.60.Pq, 14.60.Lm, 14.60.St}

\maketitle

The MiniBooNE experiment
\cite{1007.1150}
measured recently a signal
of $\bar\nu_{\mu}\to\bar\nu_{e}$ transitions
at the same ratio of distance ($L$) and energy ($E$) of that observed in the LSND experiment
\cite{hep-ex/0104049}.
This is a strong indication in favor
short-baseline $\bar\nu_{\mu}\to\bar\nu_{e}$ oscillations,
which depend just on the ratio $L/E$
(see Refs.~\cite{hep-ph/9812360,hep-ph/0211462,hep-ph/0310238,hep-ph/0405172,hep-ph/0506083,hep-ph/0606054,GonzalezGarcia:2007ib,Giunti-Kim-2007}).

In Ref.~\cite{1010.1395} we discussed the interpretation of the
MiniBooNE and LSND signals
in a minimal framework of short-baseline oscillations of antineutrinos
with a two-neutrino-like transition probability which depends on an effective mixing angle and an effective squared-mass difference,
such as that obtained in the case of four-neutrino mixing
(see Refs.~\cite{hep-ph/9812360,hep-ph/0405172,hep-ph/0606054,GonzalezGarcia:2007ib}).
The oscillations of antineutrinos may be different from those of neutrinos
\cite{hep-ph/0010178},
since the MiniBooNE experiment with a neutrino beam
did not observe a signal of short-baseline
$\nu_{\mu}\to\nu_{e}$ oscillations
\cite{0812.2243}
compatible with the MiniBooNE and LSND measurements of
$\bar\nu_{\mu}\to\bar\nu_{e}$ oscillations.
Other hints in favor of CPT-violating
different values of the effective squared-mass differences
and mixings of neutrinos and antineutrinos come from the comparison of
the data on long-baseline $\nu_{\mu}$ and $\bar\nu_{\mu}$ disappearance
in the MINOS experiment \cite{MINOS-Neutrino2010}
and from a neutrino oscillation analysis
\cite{1008.4750}
of the electron neutrino data of the Gallium radioactive source
GALLEX
\cite{1001.2731}
and
SAGE
\cite{0901.2200}
experiments
and
the electron antineutrino data of the reactor Bugey-3 \cite{Declais:1995su} and Chooz \cite{hep-ex/0301017} experiments.
Moreover,
if only antineutrino oscillation data are considered,
the strong tension between the data
of short-baseline
appearance and disappearance experiments
in 3+1 \cite{hep-ph/9607372,hep-ph/9903454,hep-ph/0405172}
and 3+2 \cite{hep-ph/0305255,0906.1997}
mixing schemes
is relaxed
\cite{hep-ph/0308299}, because the crucial data of the
CDHSW experiment \cite{Dydak:1984zq}
constrain only short-baseline $\nu_{\mu}$ disappearance
and the strong constraint coming from Super-Kamiokande atmospheric neutrino data
has been evaluated assuming equal disappearance of
$\nu_{\mu}$ and $\bar\nu_{\mu}$.

In Ref.~\cite{1010.1395} we considered the constraints
on short-baseline
$\bar\nu_{\mu}\to\bar\nu_{e}$ oscillations
coming from the data of
the KARMEN experiment \cite{hep-ex/0203021}
and the data of
the Bugey-3 \cite{Declais:1995su} and Chooz \cite{hep-ex/0301017} experiments.
The KARMEN experiment \cite{hep-ex/0203021}
did not observe short-baseline
$\bar\nu_{\mu}\to\bar\nu_{e}$ oscillations
at a distance which was about half that of LSND,
with the same neutrino energy spectrum.
Hence, the KARMEN data
constrain the parameter space of neutrino mixing
which can explain the LSND and MiniBooNE signals.
The data of the Bugey-3 \cite{Declais:1995su} and Chooz \cite{hep-ex/0301017} experiments
provide the most stringent constraints on short-baseline disappearance
of reactor $\bar\nu_{e}$'s.
For simplicity,
we considered the case in which the probability of
$\bar\nu_{e}$ disappearance
is equal to the probability of
$\bar\nu_{\mu}\to\bar\nu_{e}$ oscillations,
$P_{\bar\nu_{e}\to\bar\nu_{e}} = 1 - P_{\bar\nu_{\mu}\to\bar\nu_{e}}$.
This is the limit of the model-independent inequality
$
P_{\bar\nu_{\mu}\to\bar\nu_{e}}
\leq
1 - P_{\bar\nu_{e}\to\bar\nu_{e}}
$
which follows from simple particle conservation.

In this paper we improve the calculations presented in Ref.~\cite{1010.1395}
by considering the constraints on the mixing of $\bar\nu_{\mu}$
following from the observation of long-baseline $\bar\nu_{\mu}$ disappearance in the
MINOS experiment \cite{MINOS-Neutrino2010}.
In principle, there could be also a constraint coming from the data of
the Super-Kamiokande atmospheric neutrino experiment
\cite{1002.3471},
but since the Super-Kamiokande detector cannot distinguish neutrinos from antineutrinos
the extraction of such a constraint would require a detailed analysis of
Super-Kamiokande atmospheric neutrino data which is beyond our possibilities.
As we will see in the following,
the MINOS measurement of long-baseline $\bar\nu_{\mu}$ disappearance
is sufficient to obtain a significant constraint on the mixing of $\bar\nu_{\mu}$
which allows us to infer interesting predictions on the
short-baseline disappearance of $\bar\nu_{\mu}$'s.

The MINOS constraints on the mixing of $\bar\nu_{\mu}$
can be quantified only by considering
a specific neutrino mixing scheme.
Here,
we adopt the simplest 3+1 four-neutrino mixing scheme
(see Refs.~\cite{hep-ph/9812360,hep-ph/0405172,hep-ph/0606054,GonzalezGarcia:2007ib})
of antineutrinos in which there are three independent squared-mass differences:
\begin{enumerate}
\item
$\Delta{m}^2_{21}$ which generates the very-long-baseline disappearance of $\bar\nu_{e}$
observed by the KamLAND reactor experiment \cite{0801.4589}.
\item
$\Delta{m}^2_{31}$ which generates the long-baseline disappearance of $\bar\nu_{\mu}$
observed by the MINOS accelerator experiment \cite{MINOS-Neutrino2010}
and the oscillations of atmospheric $\bar\nu_{\mu}$'s.
\item
$\Delta{m}^2_{41}$ which generates the short-baseline $\bar\nu_{\mu}\to\bar\nu_{e}$ oscillations
observed by the LSND \cite{hep-ex/0104049} and MiniBooNE \cite{1007.1150} accelerator experiments.
\end{enumerate}
In this scheme the effective transition and disappearance probabilities in short-baseline experiments
are given by
\begin{align}
P_{\bar\nu_{\alpha}\to\bar\nu_{\beta}}^{\text{SBL}}
=
\null & \null
\sin^2 2\vartheta_{\alpha\beta}
\sin^2\left( \frac{\Delta{m}^2 L}{4 E} \right)
\,,
\label{tra}
\\
P_{\bar\nu_{\alpha}\to\bar\nu_{\alpha}}^{\text{SBL}}
=
\null & \null
1
-
\sin^2 2\vartheta_{\alpha\alpha}
\sin^2\left( \frac{\Delta{m}^2 L}{4 E} \right)
\,,
\label{sur}
\end{align}
with $\alpha\neq\beta$
and $\Delta{m}^2 = \Delta{m}^2_{41}$ for simplicity.
The effective mixing angles are related to the elements of the $4\times4$ mixing matrix $U$ of antineutrinos by
\begin{align}
\sin^2 2\vartheta_{\alpha\beta}
=
\null & \null
\sin^2 2\vartheta_{\beta\alpha}
=
4 |U_{\alpha4}|^2 |U_{\beta4}|^2
\,,
\label{mix-tra}
\\
\sin^2 2\vartheta_{\alpha\alpha}
=
\null & \null
4 |U_{\alpha4}|^2 \left( 1 - |U_{\alpha4}|^2 \right)
\,.
\label{mix-sur}
\end{align}

In this paper we consider the following data sets:
\begin{enumerate}
\renewcommand{\labelenumi}{(\theenumi)}
\renewcommand{\theenumi}{\Alph{enumi}}
\item
The LSND \cite{hep-ex/0104049},
MiniBooNE \cite{1007.1150}
and
KARMEN \cite{hep-ex/0203021}
data on short-baseline
$\bar\nu_{\mu}\to\bar\nu_{e}$
oscillations,
which depend on the product of
$|U_{e4}|^2$ and $|U_{\mu4}|^2$
through $\sin^2 2\vartheta_{e\mu}$.
We analyze the LSND and KARMEN data with the method described in Ref.~\cite{1010.1395}.
We update the analysis of MiniBooNE data presented in Ref.~\cite{1010.1395}
by using the information in the official MiniBooNE data release \cite{AguilarArevalo:2010wv-dr}.

\item
The Bugey-3 \cite{Declais:1995su} and Chooz \cite{hep-ex/0301017}
data on short-baseline $\bar\nu_{e}$ disappearance,
which depends on $|U_{e4}|^2$
through $\sin^2 2\vartheta_{ee}$.
We analyze these data with the method described in Ref.~\cite{0711.4222},
taking into account that the Chooz ratio of observed events divided by the number of expected events
in absence of oscillations must be decreased from
$R_{\text{Chooz}} = 1.010 \pm 0.028 \pm 0.036$
to
$R_{\text{Chooz}} = 0.997 \pm 0.028 \pm 0.036$
in order to remove the renormalization of the reactor $\bar\nu_{e}$ flux
done by the Chooz collaboration on the basis of the Bugey-4 integral measurement
\cite{Declais:1994ma}.

\item
The MINOS \cite{MINOS-Neutrino2010}
data on long-baseline $\bar\nu_{\mu}$ disappearance,
which constrains $|U_{\mu4}|^2$ through the inequality
\cite{hep-ph/9607372,hep-ph/9903454}
\begin{equation}
|U_{\mu4}|^4 \leq P_{\bar\nu_{\mu}\to\bar\nu_{\mu}}^{\text{MINOS}}
\,.
\label{um4}
\end{equation}
The MINOS experiment observed 97 $\bar\nu_{\mu}$ events with an expectation of
155 events in the case of no oscillations.
The corresponding integral probability of $\bar\nu_{\mu}$ survival is
\begin{equation}
P_{\bar\nu_{\mu}\to\bar\nu_{\mu}}^{\text{MINOS}}
=
0.63 \pm 0.06
\,.
\label{minos}
\end{equation}
In our analysis we constrain the value of $|U_{\mu4}|^2$
by adding to the global $\chi^2$ the MINOS contribution
\begin{equation}
\chi^2_{\text{MINOS}}
=
\left(
\frac
{\text{max}\left[0, \left(|U_{\mu4}|^4 - \overline{P}_{\bar\nu_{\mu}\to\bar\nu_{\mu}}^{\text{MINOS}}\right)\right]}
{\Delta{P}_{\bar\nu_{\mu}\to\bar\nu_{\mu}}^{\text{MINOS}}}
\right)^2
\,,
\label{chi-minos}
\end{equation}
with $\overline{P}_{\bar\nu_{\mu}\to\bar\nu_{\mu}}^{\text{MINOS}} = 0.63$
and
$\Delta{P}_{\bar\nu_{\mu}\to\bar\nu_{\mu}}^{\text{MINOS}} = 0.06$.
A more precise analysis of the MINOS energy spectrum of $\bar\nu_{\mu}$ events
taking into account the effect of $|U_{\mu4}|^2$
will be presented elsewhere
\cite{Giunti-Laveder-IP-11}.

\item
The KamLAND measurement of very-long-baseline disappearance of $\bar\nu_{e}$,
with survival probability \cite{0801.4589}
\begin{equation}
P_{\bar\nu_{e}\to\bar\nu_{e}}^{\text{KL}} = 0.61 \pm 0.03
\,.
\label{kamland}
\end{equation}
Large values of $|U_{e4}|^2$ are constrained by the inequality
\cite{hep-ph/9607372}
\begin{equation}
|U_{e4}|^4 \leq P_{\bar\nu_{e}\to\bar\nu_{e}}^{\text{KL}}
\,.
\label{ue4}
\end{equation}
In our analysis we add to the global $\chi^2$ the KamLAND contribution
\begin{equation}
\chi^2_{\text{KL}}
=
\left(
\frac
{\text{max}\left[0, \left(|U_{e4}|^4 - \overline{P}_{\bar\nu_{e}\to\bar\nu_{e}}^{\text{KL}}\right)\right]}
{\Delta{P}_{\bar\nu_{e}\to\bar\nu_{e}}^{\text{KL}}}
\right)^2
\,,
\label{chi-kamland}
\end{equation}
with $\overline{P}_{\bar\nu_{e}\to\bar\nu_{e}}^{\text{KL}} = 0.61$
and
$\Delta{P}_{\bar\nu_{e}\to\bar\nu_{e}}^{\text{KL}} = 0.03$.

\end{enumerate}

We minimized the global $\chi^2$ with respect to the three mixing parameters
$\Delta{m}^2$,
$|U_{e4}|^2$,
$|U_{\mu4}|^2$,
for which we obtained the best-fit values
\begin{equation}
\Delta{m}^2_{\text{bf}} = 0.45 \, \text{eV}^2
\,,
\
|U_{e4}|^2_{\text{bf}} = 0.0042
\,,
\
|U_{\mu4}|^2_{\text{bf}} = 0.79
\,,
\label{bf}
\end{equation}
for
\begin{equation}
\chi^2_{\text{min}} = 82.0
\,,
\quad
\text{NDF} = 83
\,,
\quad
\text{GoF} = 51\%
\,,
\label{chimin}
\end{equation}
where
NDF is the number of degrees of freedom
and
GoF is the goodness-of-fit.
Hence the global fit is acceptable.
Moreover,
the parameter goodness-of-fit
\cite{hep-ph/0304176}
is 28\%,
which is reasonable.

\begin{figure*}[t!]
\begin{center}
\includegraphics*[bb=5 11 571 571, width=0.8\linewidth]{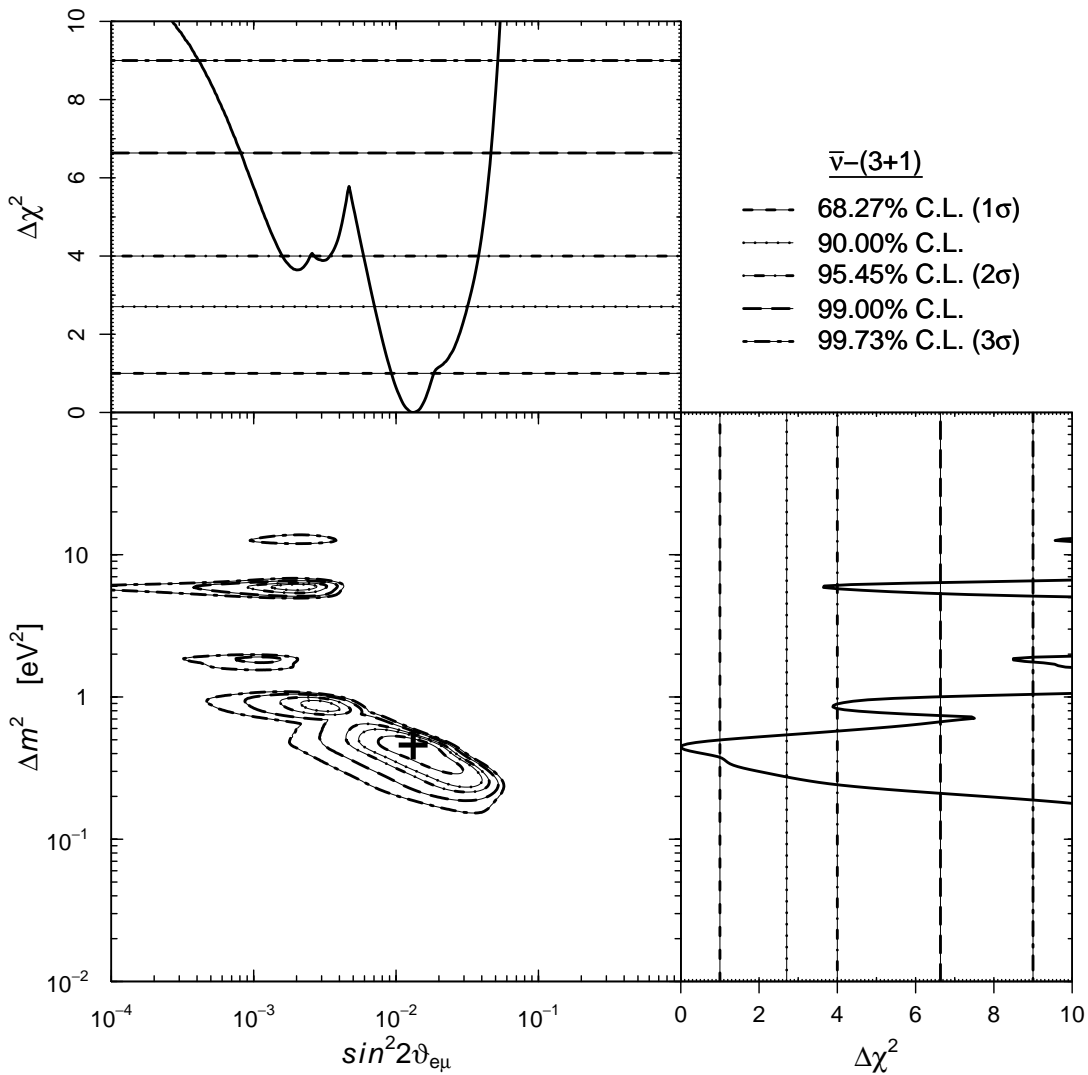}
\end{center}
\caption{ \label{elmu}
Allowed regions in the
$\sin^2 2\vartheta_{e\mu}$--$\Delta{m}^{2}$ plane
and
marginal $\Delta\chi^{2}=\chi^2-\chi^2_{\text{min}}$'s
for
$\sin^2 2\vartheta_{e\mu}$ and $\Delta{m}^{2}$.
The best-fit point is indicated by a cross.
}
\end{figure*}

\begin{figure*}[t!]
\begin{center}
\includegraphics*[bb=5 11 565 571, width=0.8\linewidth]{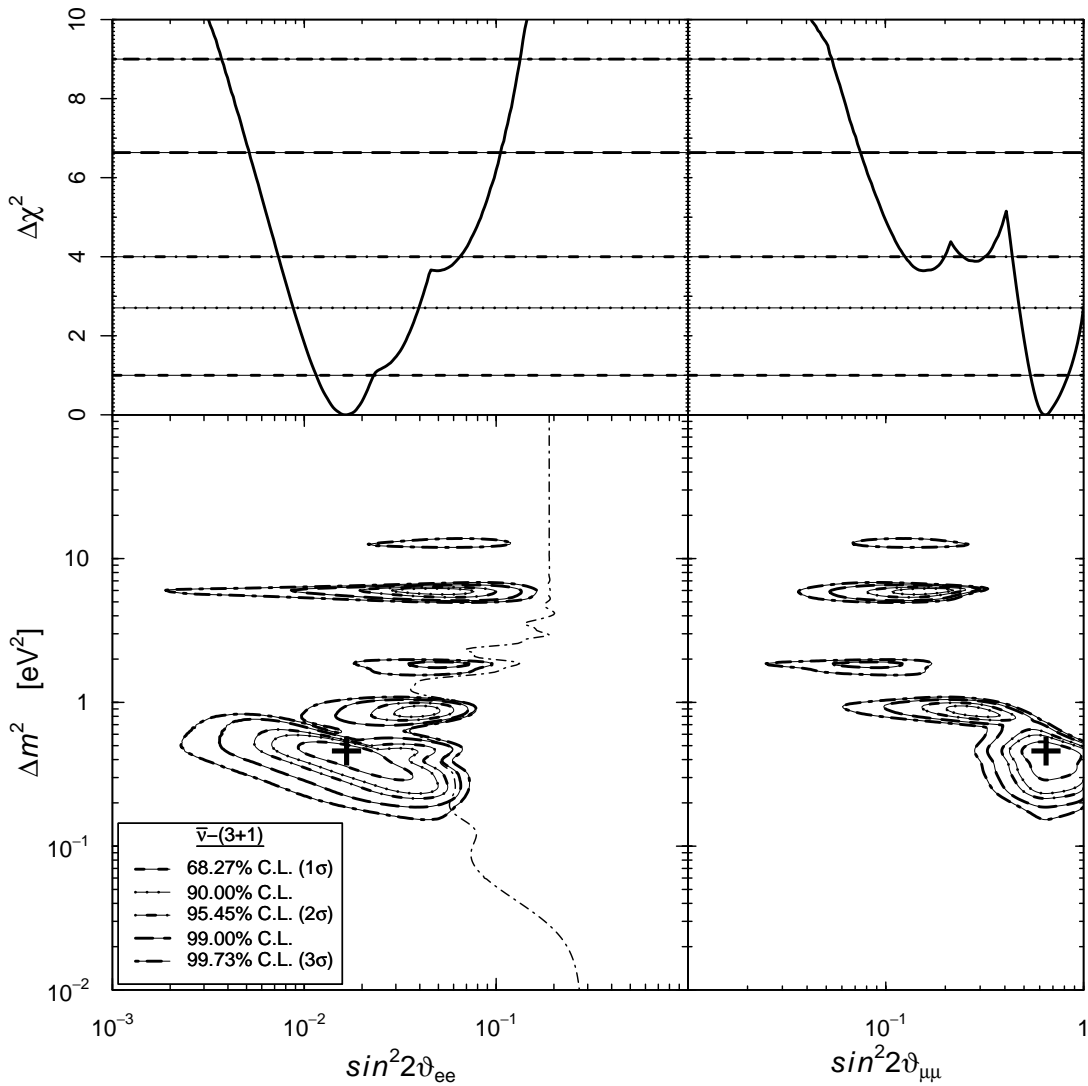}
\end{center}
\caption{ \label{see-smm}
Allowed regions in the
$\sin^2 2\vartheta_{ee}$--$\Delta{m}^{2}$
and
$\sin^2 2\vartheta_{\mu\mu}$--$\Delta{m}^{2}$
planes
and
marginal $\Delta\chi^{2}=\chi^2-\chi^2_{\text{min}}$'s
for
$\sin^2 2\vartheta_{ee}$ and $\sin^2 2\vartheta_{\mu\mu}$.
The best-fit point is indicated by a cross.
The thin dash-dotted line in the
$\sin^2 2\vartheta_{ee}$--$\Delta{m}^{2}$
plane
represents the $3\sigma$ exclusion curve obtained from
the reactor Bugey-3 and Chooz data.
}
\end{figure*}

The best-fit values of the effective oscillation amplitudes corresponding to
$|U_{e4}|^2_{\text{bf}}$
and
$|U_{\mu4}|^2_{\text{bf}}$
in Eq.~(\ref{bf})
are
\begin{equation}
\sin^2 2\vartheta_{e\mu}^{\text{bf}} = 0.013
\,,
\
\sin^2 2\vartheta_{ee}^{\text{bf}} = 0.017
\,,
\
\sin^2 2\vartheta_{\mu\mu}^{\text{bf}} = 0.65
\,,
\label{sinbf}
\end{equation}
The allowed regions in the
$\sin^2 2\vartheta_{e\mu}$--$\Delta{m}^{2}$,
$\sin^2 2\vartheta_{ee}$--$\Delta{m}^{2}$ and
$\sin^2 2\vartheta_{\mu\mu}$--$\Delta{m}^{2}$ planes
are shown
in Figs.~\ref{elmu} and \ref{see-smm},
together with the marginal $\Delta\chi^{2}=\chi^2-\chi^2_{\text{min}}$'s
for
$\Delta{m}^{2}$,
$\sin^2 2\vartheta_{e\mu}$,
$\sin^2 2\vartheta_{ee}$ and
$\sin^2 2\vartheta_{\mu\mu}$.

Figure~\ref{elmu} is similar to Fig.~7 of Ref.~\cite{1010.1395}.
In Fig.~\ref{elmu} the constraint on $|U_{\mu4}|^2$ from MINOS data
shifts the allowed interval of $\sin^2 2\vartheta_{e\mu}=4 |U_{e4}|^2 |U_{\mu4}|^2$
towards slightly smaller values with respect to those in Fig.~7 of Ref.~\cite{1010.1395},
where the upper bounds on $\sin^2 2\vartheta_{e\mu}$ are given only by the reactor
constraints on $|U_{e4}|^2$, allowing $|U_{\mu4}|^2$ to be as large as unity.
However, the change in the allowed intervals of
$\Delta{m}^{2}$ and $\sin^2 2\vartheta_{e\mu}$
with respect to those obtained in Ref.~\cite{1010.1395}
is rather small:
from the marginal $\Delta\chi^{2}$'s in Fig.~\ref{elmu}
we obtain
\begin{eqnarray}
&
2\times10^{-3} \lesssim \sin^2 2\vartheta_{e\mu} \lesssim 4\times10^{-2}
\,,
&
\label{sem}
\\
&
0.2 \lesssim \Delta{m}^2 \lesssim 1 \, \text{eV}^2
\quad
\text{or}
\quad
\Delta{m}^2 \simeq 6 \, \text{eV}^2
\,,
&
\label{dm2}
\end{eqnarray}
at
95\% C.L.
(to be compared with
$2\times10^{-3} \lesssim \sin^2 2\vartheta_{e\mu} \lesssim 5\times10^{-2}$
and
$0.2 \lesssim \Delta{m}^2 \lesssim 2 \, \text{eV}^2$
obtained in Ref.~\cite{1010.1395}).

Figure~\ref{see-smm} shows the allowed regions in the
$\sin^2 2\vartheta_{ee}$--$\Delta{m}^{2}$ plane,
together with the $3\sigma$ exclusion curve obtained from
the reactor Bugey-3 and Chooz data.
One can see that $\sin^2 2\vartheta_{ee}$
is approximately bounded to be smaller than the limit imposed by the reactor data.
Taking into account the approximation
\begin{equation}
\sin^2 2\vartheta_{ee} \simeq 4 |U_{e4}|^2
\,,
\label{see-ue4}
\end{equation}
which is valid for the small values of $|U_{e4}|^2$ allowed by KamLAND data
(Eq.~(\ref{ue4})),
the lower limits on $\sin^2 2\vartheta_{ee}$
follow from the need to have a value of $\sin^2 2\vartheta_{e\mu} = 4 |U_{e4}|^2 |U_{\mu4}|^2$
in the range in Eq.~(\ref{sem}) with $|U_{\mu4}|^2$ limited to be smaller than unity
by $\chi^2_{\text{MINOS}}$ in Eq.~(\ref{chi-minos}).
From the marginal $\Delta\chi^{2}$ in Fig.~\ref{see-smm}
we obtain
\begin{equation}
7\times10^{-3} \lesssim \sin^2 2\vartheta_{ee} \lesssim 6\times10^{-2}
\,,
\label{see}
\end{equation}
at
95\% C.L..

Figure~\ref{see-smm} shows also the allowed regions in the
$\sin^2 2\vartheta_{\mu\mu}$--$\Delta{m}^{2}$ plane
and the marginal $\Delta\chi^{2}$ for $\sin^2 2\vartheta_{\mu\mu}$,
which gives
\begin{equation}
\sin^2 2\vartheta_{\mu\mu} \gtrsim 0.1
\,,
\label{smm}
\end{equation}
at
95\% C.L..
This result is interesting, because it implies that
the short-baseline disappearance of $\bar\nu_{\mu}$'s
is rather large and could be measured in future experiments
\cite{0909.0355,0910.2698,AndreRubbia:NEU2012}.
The preferred region in Fig.~\ref{see-smm} lies around the best-fit point
in Eq.~(\ref{sinbf}) which corresponds to a rather large value of
$\sin^2 2\vartheta_{\mu\mu}$.
Notice that such large values of
$\sin^2 2\vartheta_{\mu\mu}$
are not constrained by MINOS data,
because they correspond to values of $|U_{\mu4}|^2$
close to 1/2.
MINOS data constrain small values of
$\sin^2 2\vartheta_{\mu\mu} = 4 |U_{\mu4}|^2 \left( 1 - |U_{\mu4}|^2 \right)$
in conjunction with the need to have a value of $\sin^2 2\vartheta_{e\mu} = 4 |U_{e4}|^2 |U_{\mu4}|^2$
in the range in Eq.~(\ref{sem}) with
a small $|U_{e4}|^2 \simeq \sin^2 2\vartheta_{ee} / 4$
from Eq.~(\ref{see}).

It is interesting to notice that in Fig.~\ref{see-smm}
large values of
$\sin^2 2\vartheta_{\mu\mu}$
are excluded for
$\Delta{m}^2 \gtrsim 1 \, \text{eV}^2$
by the constraints imposed by 
MiniBooNE $\bar\nu_{\mu}$ data,
which are included in the analysis
according to the method described in Ref.~\cite{1010.1395}
taking into account the $\bar\nu_{\mu}$ disappearance
given by Eq.~(\ref{sur}).
This is in agreement with the MiniBooNE exclusion curve
for $\bar\nu_{\mu}$ disappearance in Fig.~3 of Ref.~\cite{0903.2465}.

\begin{figure*}[t!]
\begin{center}
\includegraphics*[bb=7 14 505 385, width=0.8\linewidth]{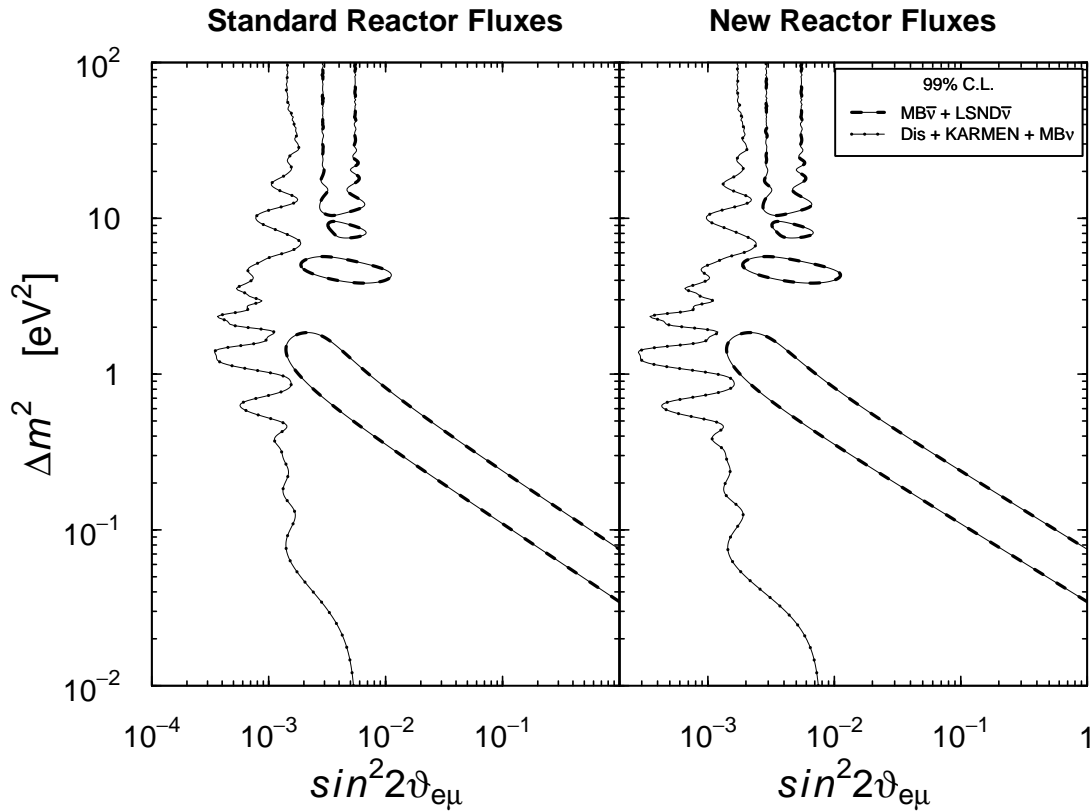}
\end{center}
\caption{ \label{sup2}
Comparison of the
regions in the
$\sin^2 2\vartheta_{e\mu}$--$\Delta{m}^{2}$
plane allowed at 99\% C.L. by
LSND \cite{hep-ex/0104049} and MiniBooNE \cite{1007.1150}
$\bar\nu_{\mu}\to\bar\nu_{e}$ data on the right of each panel
with the 99\% C.L.
exclusion curve on the left of each panel obtained from
MiniBooNE $\nu_{\mu}\to\nu_{e}$ data \cite{0812.2243},
KARMEN $\bar\nu_{\mu}\to\bar\nu_{e}$ data \cite{hep-ex/0203021},
CDHSW $\nu_{\mu}\to\nu_{\mu}$ data \cite{Dydak:1984zq}
atmospheric neutrino data \cite{0705.0107}
and
Bugey-3 \cite{Declais:1995su} and Chooz \cite{hep-ex/0301017} $\bar\nu_{e}\to\bar\nu_{e}$ data
with the standard reactor $\bar\nu_{e}$ fluxes (left panel)
and
the new reactor $\bar\nu_{e}$ fluxes
of Ref.~\cite{1101.2663,1101.2755}
(right panel).
}
\end{figure*}

In conclusion,
we have analyzed the data of
short-baseline antineutrino oscillation experiments
taking into account the constraints on the mixing of $\bar\nu_{\mu}$
given by the observation of long-baseline $\bar\nu_{\mu}$ disappearance in the
MINOS experiment \cite{MINOS-Neutrino2010}
in the framework of 3+1 antineutrino mixing.
The LSND \cite{hep-ex/0104049} and MiniBooNE \cite{1007.1150} signals in favor of
short-baseline
$\bar\nu_{\mu}\to\bar\nu_{e}$ oscillations
are compatible with the constraints given by the data of
the KARMEN \cite{hep-ex/0203021} short-baseline $\bar\nu_{\mu}\to\bar\nu_{e}$ experiment,
the Bugey-3 \cite{Declais:1995su} and Chooz \cite{hep-ex/0301017}
short-baseline $\bar\nu_{e}\to\bar\nu_{e}$ experiments,
the MINOS \cite{MINOS-Neutrino2010}
long-baseline $\bar\nu_{\mu}\to\bar\nu_{\mu}$ experiments
and the KamLAND \cite{0801.4589}
very-long-baseline $\bar\nu_{e}\to\bar\nu_{e}$ experiment.
Our analysis predicts that the short-baseline disappearance of
$\bar\nu_{\mu}$
is rather large and could be measured in future short-baseline
$\bar\nu_{\mu}$ disappearance experiments sensitive
to values of
$\Delta{m}^2$
in the sub-eV$^2$ region
\cite{0909.0355,0910.2698,AndreRubbia:NEU2012}.

Although the numerical results obtained in this paper
depend on the chosen framework of 3+1 antineutrino mixing,
the prediction of large $\bar\nu_{\mu}$ disappearance in short-baseline
experiments is a general consequence of
the LSND and MiniBooNE signals in favor of
short-baseline
$\bar\nu_{\mu}\to\bar\nu_{e}$ oscillations.
In fact,
since the mixing of $\bar\nu_{e}$
with the massive neutrino(s) responsible for short-baseline oscillations
is constrained to be small by the
short-baseline reactor $\bar\nu_{e}$ data,
taking into account the KamLAND measurement of
a large very-long-baseline $\bar\nu_{e}$ disappearance,
the mixing of $\bar\nu_{\mu}$
with the massive neutrino(s) responsible for short-baseline oscillations
must be relatively large.
The MINOS measurement of
long-baseline $\bar\nu_{\mu}$ disappearance
implies that $\bar\nu_{\mu}$ must have also a relatively large mixing
with the massive neutrino(s) responsible for long-baseline oscillations.
Therefore since $\bar\nu_{\mu}$ have relatively large mixing with the two sets of massive neutrinos
whose squared-mass difference generate short-baseline oscillations,
the amplitude of short-baseline $\bar\nu_{\mu}$ disappearance
must be large.
The numerical predictions for such amplitude in
mixing schemes more complicated than the simplest
framework of 3+1 antineutrino mixing considered here
will be presented elsewhere
\cite{Giunti-Laveder-IP-11}.

\bigskip
\centerline{\textbf{Note Added}}
\medskip

After the completion of this work,
a very interesting new evaluation of the $\bar\nu_{e}$ fluxes produced in nuclear reactors
has been published in Ref.~\cite{1101.2663}.
The increase of about 3\% of the flux normalization with respect to the standard evaluation
used in the analysis of all experimental data
(see Ref.~\cite{hep-ph/0107277})
has several implications for the interpretation of neutrino oscillation data
and may lead to a reactor antineutrino anomaly
\cite{1101.2755}.
Such an increase of the reactor $\bar\nu_{e}$ fluxes tends to decrease the
tension between the putative lack of $\bar\nu_{e}$ and $\nu_{\mu}$
short-baseline disappearance and
the
LSND and MiniBooNE signals of
short-baseline $\bar\nu_{\mu}\to\bar\nu_{e}$ oscillations
in CPT-invariant 3+1 neutrino mixing schemes
\cite{hep-ph/9607372,hep-ph/9812360,hep-ph/9903454,hep-ph/0102252,hep-ph/0207157,hep-ph/0405172,0705.0107},
reducing the need to treat the oscillations of neutrinos and antineutrinos separately \cite{1010.1395}.
Figure~\ref{sup2}
illustrates the change
by comparing the
regions in the
$\sin^2 2\vartheta_{e\mu}$--$\Delta{m}^{2}$
plane allowed at 99\% C.L. by
LSND \cite{hep-ex/0104049} and MiniBooNE \cite{1007.1150}
$\bar\nu_{\mu}\to\bar\nu_{e}$ data
with the 99\% C.L.
exclusion curve obtained from
MiniBooNE $\nu_{\mu}\to\nu_{e}$ data \cite{0812.2243},
KARMEN $\bar\nu_{\mu}\to\bar\nu_{e}$ data \cite{hep-ex/0203021},
CDHSW $\nu_{\mu}\to\nu_{\mu}$ data \cite{Dydak:1984zq}
atmospheric neutrino data \cite{0705.0107}
and
Bugey-3 \cite{Declais:1995su} and Chooz \cite{hep-ex/0301017} $\bar\nu_{e}\to\bar\nu_{e}$ data
with the standard reactor $\bar\nu_{e}$ fluxes
and
the new reactor $\bar\nu_{e}$ fluxes.
One can see that the change is very small.
The parameter goodness-of-fit shifts from
0.0048\%
to
0.0064\%.
Since the new reactor $\bar\nu_{e}$ fluxes do not allow us to
reconcile the data in the framework of CPT-invariant 3+1 neutrino mixing,
the analysis of the antineutrino data presented in this paper remains valid.
More detailed implications of the new reactor $\bar\nu_{e}$ fluxes
will be discussed elsewhere
\cite{Giunti-Laveder-IP-11}.

\bibliography{bibtex/nu}

\end{document}